\documentclass[9pt]{article}
\usepackage{spconf,amsmath,graphicx,hyperref}
\usepackage{multirow}
\usepackage{booktabs}
\usepackage{xcolor}
\usepackage{siunitx}
\usepackage[inline]{enumitem}
\usepackage{cleveref}
\definecolor{xiaomi_gray}{HTML}{A9A9A9}
\definecolor{catpuccin_red}{HTML}{f38ba8}
\definecolor{catpuccin_blue}{HTML}{89b4fa}

\title{GLAP: General contrastive audio-text pretraining across domains and languages}
\name{\parbox{\textwidth}{\centering Heinrich Dinkel, Zhiyong Yan, Tianzi Wang, Yongqing Wang, Xingwei Sun, Yadong Niu, Jizhong Liu, Gang Li, Junbo Zhang, Jian Luan}}
\address{MiLM Plus, Xiaomi Inc., China}
\begin{document}

%\ninept
%
\maketitle
\begin{abstract}
Contrastive Language Audio Pretraining (CLAP) is a widely-used method to bridge the gap between audio and text domains.
Current CLAP methods enable sound and music retrieval in English, ignoring multilingual spoken content. 
To address this, we introduce general language audio pretraining (GLAP), which expands CLAP with multilingual and multi-domain abilities.
GLAP demonstrates its versatility by achieving competitive performance on standard audio-text retrieval benchmarks like Clotho and AudioCaps, while significantly surpassing existing methods in speech retrieval and classification tasks.
Additionally, GLAP achieves strong results on widely used sound-event zero-shot benchmarks, while simultaneously outperforming previous methods on speech content benchmarks.
Further keyword spotting evaluations across 50 languages emphasize GLAP's advanced multilingual capabilities. 
Finally, multilingual sound and music understanding is evaluated across four languages.
\end{abstract}
\section{Introduction}

\begin{figure}[tb]
    \centering
    \includegraphics[width=1.0\linewidth]{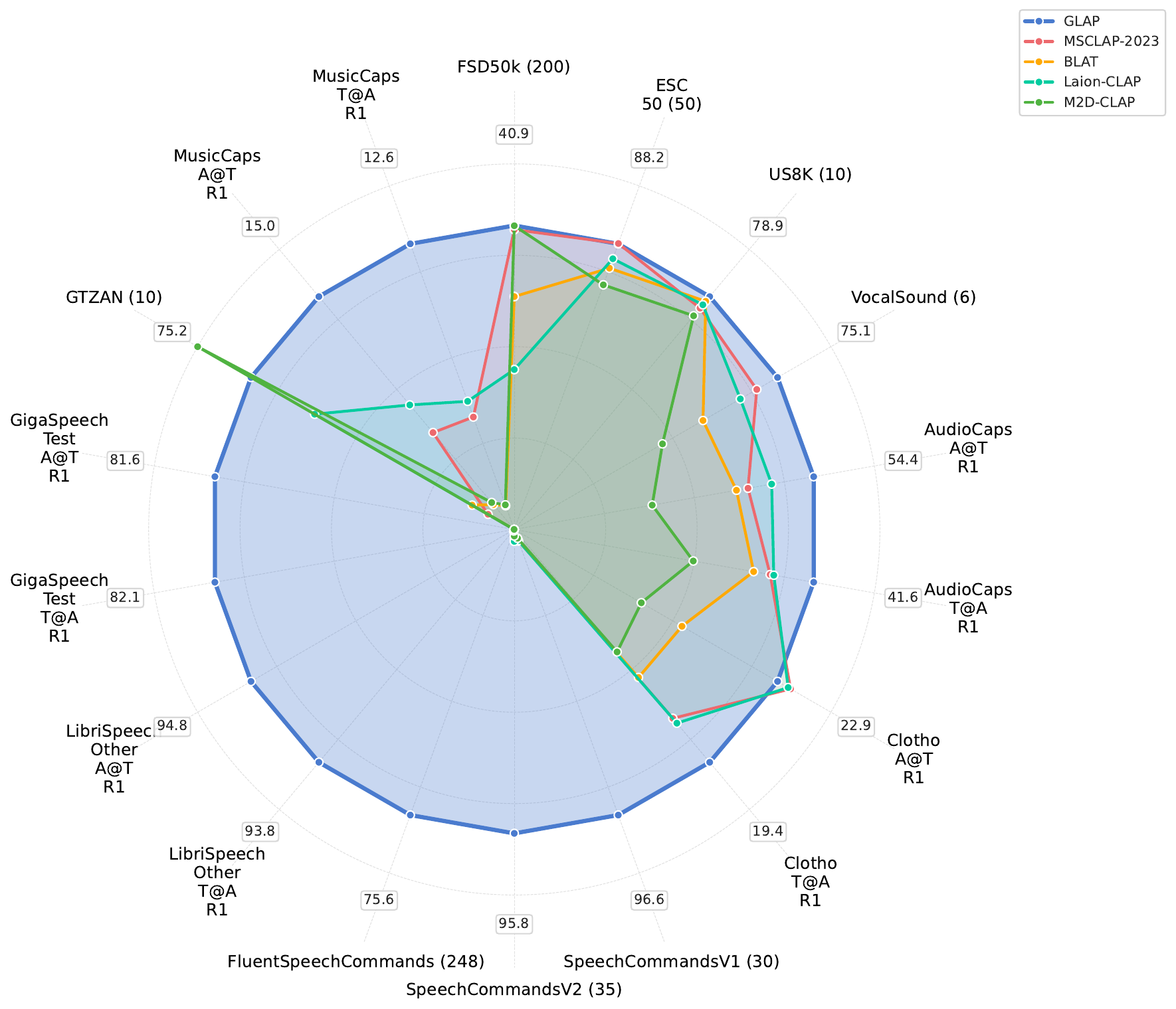}
    \caption{GLAP’s retrieval and zero-shot performance. A@T and T@A represent retrieval tasks {of Audio-to-Text and Text-to-Audio, respectively}, others are zero-shot (\#labels).
    }
    \label{fig:glap_capabilities}
\end{figure}

In the field of computer vision, Contrastive Language-Image Pretraining (CLIP)~\cite{Radford2021_CLIP} represents a significant breakthrough in extracting efficient representations that can be applied across various downstream tasks and domains.
Similarly, Contrastive Language-Audio Pretraining (CLAP)~\cite{laionclap2023,xu2024blat} bridges text and audio, enabling zero-shot transfer learning.
However, existing CLAP models, including those trained on large datasets like the one in~\cite{elizalde2024natural} and multilingual extensions~\cite{yan24_interspeech}, perform poorly on basic speech tasks like keyword spotting. 
% While these models excel with sound and music, they lack comprehensive speech understanding.
CLAP embeddings primarily target sound and music, missing comprehensive speech representation (i.e., spoken language) - a critical aspect of audio processing.
This work proposes general language audio pretraining (GLAP) aimed at aligning speech content with text, without compromising in sound and music performance. 
GLAP further effectively generalizes its speech and sound understanding capabilities beyond English which be seen in \Cref{fig:glap_capabilities}.

\section{General language audio pretraining}

To enable speech understanding in CLAP models, which are trained on sound and music data, one simple solution is to add speech data to the training dataset.
However, as our analysis in \Cref{ssec:audio_encoder} shows, this approach leads to compromised performance due to the absence of a unified audio encoder. 
Thus GLAP's main goal is to provide a unified framework framework that maintains high performance for sound, music and speech retrieval tasks, enabling alignment across these audio modalities.
GLAP is trained with pairs of audio-text samples $(a,t)$ in contrastive fashion.
Features from these audio-text pairs are extracted through a pre-trained multi-lingual text encoder ${\text{E}_{T}}$ and a pre-trained general audio encoder ${\text{E}_A}$:
\begin{equation*}
\begin{aligned}
e_a&=\text{MLP}_{A}(\text{E}_A(a)), & e_t =\text{MLP}_{T}(\text{E}_{T}(t)),\\
\end{aligned}
\end{equation*}
A trainable multi-layer perceptron (MLP) is added to align the dimensions. 
Finally, the pair $(e_a, e_t)$ is scored using cosine distance $s = \frac{e_a\cdot e_t^T} {||e_a|| \cdot ||e_t||}$.
Unlike previous works, we use the sigmoid loss~\cite{zhai2023sigmoid} as our main training objective $\mathcal{L}$, computed as:
% \vspace{-1em}
\begin{align}
\mathcal{L} &= -\frac{1}{B} \sum_{i}^{B} \sum_{j}^{B} \log \sigma\left(s'(i,j) \cdot \psi[i, j]\right), \\
s'(i,j) &= \frac{s(i,j) + \beta}{\tau},
\label{eq:siglip}
\end{align}

where $\sigma$ is the sigmoid function, $B$ is the batchsize, $\beta,\tau$ are learnable parameters, ``$\cdot$'' is the element-wise product, and 

$$\psi[i, j] =
\begin{cases} 
1 & \text{if } i = j, \\
-1 & \text{otherwise}.
\end{cases}$$

\begin{figure}
    \centering
    \includegraphics[width=0.98\linewidth]{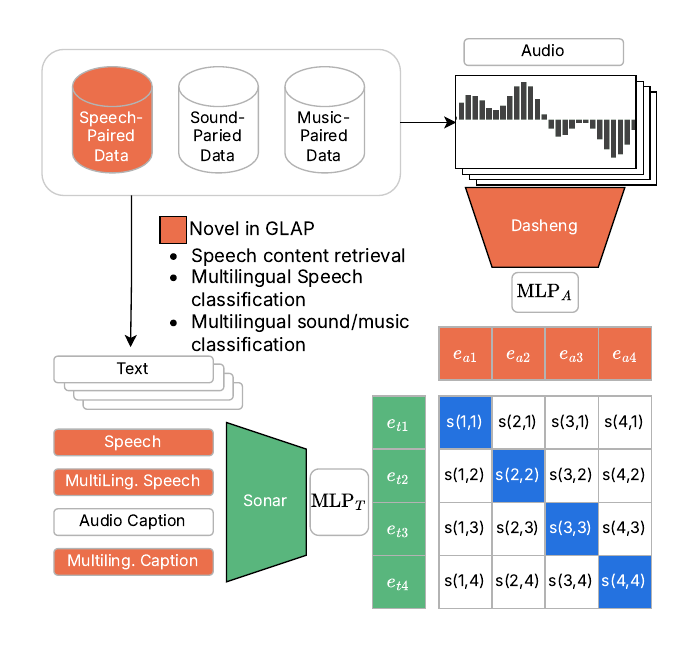}
    \caption{GLAP enables multilingual speech-content retrieval, ontop of the standard sound/music capabilities.} 
    \label{fig:glap_framework}
\end{figure}

The primary reason for choosing sigmoid loss over standard cross-entropy is its superior performance with large batch sizes and datasets, as we observed performance boosts of 1\% to 5\% across all retrieval tasks.
An overview of the proposed framework can be seen in \Cref{fig:glap_framework}.

% MLCLAP https://arxiv.org/pdf/2310.11830
\section{Experimental Setup}
\label{sec:exp}

All audio data is preprocessed by resampling all datasets to a single channel at 16 kHz. 
Following~\cite{yan24_interspeech}, we use the text encoder from Sonar~\cite{Duquenne:2023:sonar_arxiv} for text embeddings.
We initialize the loss parameters as $\tau=0.07,\beta=-10$ (\Cref{eq:siglip}).
We train the model with an effective batch size of $B=1024$ (128 per GPU) for a maximum of 20 epochs (10,000 batches per epoch). 
Training uses an 8-bit Adam optimizer with a cosine decay scheduler, where the learning rate warms up from 0 to ${10^{-4}}$ over the first two epochs and then decays to ${10^{-5}}$.
The source code and checkpoints are publicly available\footnote{\url{github.com/xiaomi-research/dasheng-glap}}.

% We used an effective batch-size $B=1024$ across 8 A800 GPU's, meaning that each GPU used a batch-size of $128$.
\begin{table}[ht]
    \centering
    % \footnotesize
    \begin{tabular}{rr|lll}
    \toprule
        Domain & Dataset &  hours & \#Pairs & \#Lang  \\
        \midrule
         \multirow{4}{*}{Speech}  & YODAS~\cite{li2023yodas}  & 400 k & 431 M & 145 \\
         & GigaSpeech~\cite{chen2021gigaspeech} & 10 k & 8 M  & 1  \\
         & LibriSpeech~\cite{panayotov2015librispeech} & 960 & 271 k & 1 \\
         & AISHELL-1~\cite{bu2017aishell} & 180 & 131 k & 1 \\
        \midrule
         \multirow{6}{*}{Sound}  & Sound-VECaps$_A$~\cite{yuan2024sound} &  5200 &  1.6 M & {1+7} \\
         & Auto-ACD~\cite{sun2024auto} &  5200  & 1.8 M & {1+7} \\
         & AudiosetCaps~\cite{bai2024audiosetcaps} & 5700  &  2.0 M & {1+7} \\
         & WavCaps~\cite{mei2023wavcaps} & 7544  &  400 k &  {1+7} \\
         & AudioCaps~\cite{audiocaps} & 127 & 49 k &  {1+7}\\
         & Clothov2~\cite{drossos2020clotho} & 35  & 4884 &  {1+7} \\
        \midrule
        \multirow{2}{*}{Music}  & MusicCaps~\cite{agostinelli2023musiclm_musiccaps} & 7.3 & 2640 &  {1+7} \\
         & Songdescriber~\cite{manco2023songdescriber} &  12  &  360 &  {1+7} \\
        % \midrule
        % $\sum$ &  
        \bottomrule
    \end{tabular}
    \caption{
    Training datasets with duration (hours), audio-text pairs (\# Pairs), and languages (\# Lang). Music and sound data labeled in English were auto-translated to 7 additional languages via Sonar.}
    \label{tab:dataset}
\end{table}
\begin{table*}[t]
\centering
\resizebox{\linewidth}{!}{
\begin{tabular}{r|llll|llll|llll}
\toprule
& \multicolumn{4}{c|}{LibriSpeechOther} & \multicolumn{4}{c|}{MusicCaps} & \multicolumn{4}{c}{AISHELL2-Test} \\
& \multicolumn{2}{c}{Text-to-Audio} & \multicolumn{2}{c|}{Audio-to-Text} & \multicolumn{2}{c}{Text-to-Audio} & \multicolumn{2}{c|}{Audio-to-Text} & \multicolumn{2}{c}{Text-to-Audio} & \multicolumn{2}{c}{Audio-to-Text} \\
 & R@1 & R@10 & R@1 & R@10 & R@5 & R@10 & R@5 & R@10 & R@1 & R@10 & R@1 & R@10 \\
\midrule
MSCLAP-2022$^\dagger$~\cite{elizalde2022clap} & 0.1 & 0.6 & 0.0 & 0.4 & 4.3 & 7.2 & 5.2 & 7.4 & 0.0 & 0.2 & 0 & 0.18 \\
MSCLAP-2023$^\dagger$~\cite{elizalde2024natural} & 0.1 & 0.4 & 0.1 & 0.2 & 14.4 & 21.7 & 17.7 & 25.9 & 0.1 & 0.2 & 0.0 & 0.2 \\
L-CLAP$^\dagger$~\cite{laionclap2023} & 0.1 & 0.8 & 0.1 & 0.5 & 17.2 & 25.5 & \textbf{22.0} & 31.1 & 0 & 0.2 & 0.0 & 0.2 \\
L-CLAP$_{\text{Speech-Music}}^\dagger$~\cite{laionclap2023} & 0.1 & 0.9 & 0.1 & 0.9 & 16.8 & 25.4 & 16.8 & 25.2 & 0 & 0.2 & 0.0 & 0.2 \\
% LCLAP$_{\text{Speech-Music-General}}^\dagger$~\cite{laionclap2023} & 0.07 & 0.61 & 0.1 & 0.51 & 19.13 & 27.23 & 21.69 & 31.52 & 0 & 0.1 & 0.02 & 0.22 \\
COLLAP-Roberta~\cite{wu2024collap} & - & - & - & - & 15.2 & - & 9.5 & - & - & - & - & - \\
COLLAP-GPT~\cite{wu2024collap} & - & - & - & - & 17.4 & - & 10.3 & - & - & - & - & - \\
BLAT$^\dagger$~\cite{xu2024blat} & 0.0 & 0.8 & 0 & 0.4 & 3.2 & 5.1 & 3.9 & 5.8 & 0.0 & 0.2 & 0.0 & 0.2 \\
M2D-Clap$^\dagger$~\cite{niizumi24_interspeech} & 0.1 & 0.6 & 0.0 & 0.4 &4.3& 7.2 & 5.2 & 7.4 & 0.1 & 0.2 & 0.1 & 0.3 \\
\hline
GLAP & \textbf{93.8} & \textbf{96.8} & \textbf{91.8} & \textbf{94.4} & \textbf{30.3} & \textbf{41.2} & 15.0 & \textbf{44.4} & \textbf{98.5} & \textbf{99.7} & \textbf{99.1} & \textbf{99.7} \\
\bottomrule
\end{tabular}
}
\caption{Music/Speech retrieval results. $^\dagger$ indicates evaluation from a public checkpoint. Best in bold; higher is better.}
\label{tab:retreval_speech_music}
\end{table*}
\subsection{Datasets}
\label{ssec:dataset}

\paragraph*{Training}

We trained our model on a diverse range of audio-text datasets, as detailed in \Cref{tab:dataset}. Our main dataset is YODAS~\cite{li2023yodas}, a 400k-hour YouTube corpus primarily labeled with noisy automated speech-to-text pipelines. 
To improve data quality, we supplemented YODAS with cleaner English datasets (GigaSpeech, LibriSpeech) and a Chinese dataset (AISHELL-1).
While YODAS covers 145 languages, relying solely on multilingual speech data limits generalization in sound/music retrieval. 
To address this, we followed~\cite{yan24_interspeech} and leveraged Sonar~\cite{Duquenne:2023:sonar_arxiv} to translate the original English captions of all sound and music datasets into other seven widely spoken languages: German, Chinese, Catalan, Spanish, Japanese, French, and Dutch.
To address the imbalance of speech data shown in \Cref{tab:dataset}, we categorized the data into four groups: sound + music, English speech, Chinese speech, and other languages and sample equally from each group.

\vspace{-1em}
\paragraph*{Evaluation}
We evaluate retrieval performance across sound datasets such as Auto-ACD (ACD)~\cite{sun2024auto}, AudioCaps (AC)~\cite{audiocaps}, Clothov2 (Clotho)~\cite{drossos2020clotho}, music datasets such as MusicCaps (MC)~\cite{agostinelli2023musiclm_musiccaps} and the speech datasets LibriSpeech (LS)~\cite{panayotov2015librispeech}, GigaSpeech~\cite{chen2021gigaspeech} and AISHELL-2 (AIS2)~\cite{bu2017aishell}.
For zero-shot classification, we primarily follow~\cite{elizalde2024natural} and evaluate on ESC-50, FSD50K, UrbanSound8K (US8K), CREMA-D, GTZAN, NSynth instruments, Beijing-Opera, VocalSound, as well as Speech Commands V1/V2 (SCV1/2) and Fluent Speech Commands (FSC)~\cite{lugosch2019speech}.
All test datasets with the exception of AIS2 are labeled in English.

\subsection{Evaluation metrics and Zero-shot Prompting}
\label{ssec:eval_metrics}

Audio-text retrieval performance is measured using recall at rank (R@k), which checks if the correct item is in the top k results. 
For a more complete comparison, we also use mean average precision at rank 10 (mAP10). 
For zero-shot inference, we use accuracy for single-class classification and mean average precision (mAP) for multi-label classification, consistent with standard practice~\cite{elizalde2024natural}.
GLAP performs zero-shot inference, but without a dedicated token to differentiate between spoken words and sound events, the model relies on effective text prompts. Prompts used in this work are shown in \Cref{tab:prompts}.

% \subsection{Prompting}
% \label{ssec:prompts}

\begin{table}[t]

    \centering
    % \footnotesize
    \begin{tabular}{r|l}
    \toprule
        Task & Prompt  \\
    \midrule
         Speech & \{label\}\\
         Music & The music in the style of \{label\}. \\
         Sound & The sound of \{label\} can be heard. \\
         \bottomrule
    \end{tabular}
    \caption{Prompts for zero-shot evaluation.}
    \label{tab:prompts}
\end{table}

\section{Results}
\label{sec:results}

% In this section we present out results. 

\subsection{Audioencoder investigation}
\label{ssec:audio_encoder}

In our view, a key limitation lies in the reliance on sound-event audio encoders for CLAP.
For this purpose, we compare the proposed training framework using five distinct audio-encoders, being Dasheng~\cite{dinkel24b_interspeech}, CED-Base~\cite{dinkel2023ced}, Beats~\cite{chen2022beats}, Whisper-Base~\cite{radford2023robust} and WavLM~\cite{chen2022wavlm}, each with $\approx$ 90M parameters and support for variable length inputs.
Each encoder is initialized from a publicly available checkpoint and trained using the settings in \Cref{sec:exp}.
As the results in \Cref{tab:audio_encoder} indicate, sound-event encoders like CED and Beats excel at music and sound tasks, but perform poorly on speech. 
Conversely, Whisper and WavLM are strong on speech but weak on other audio types. 
Dasheng is the most versatile, performing well across all domains. 
Because of this, further experiments use Dasheng as the primary audio encoder.

\begin{table}[tb]
    \centering
    \begin{tabular}{l|rr|r|rr}
    \toprule
       \multirow{2}{*}{Encoder} & \multicolumn{2}{c|}{Sound} & Music & \multicolumn{2}{c}{Speech} \\       
          &  AC & ACD & MC & LS-other & AIS2 \\
    \midrule         
         CED-Base & 58.6  & 62.0  & 25.1 & 87.8  & 70.6 \\
         Beats &  55.1 &  64.3  & 23.9 &  91.8 & 44.0  \\
         Whisper-Base & 46.5 &  52.9 & 15.8 & 98.9 & 99.4 \\
         WavLM & 36.1  & 47.5  & 14.8 & 99.9  & 96.3 \\
         \hline
         Dasheng & 55.8 & 60.1 & 20.3 & 94.8 & 99.0 \\
    \bottomrule
    \end{tabular}
    \caption{Text-to-Audio retrieval performance across five datasets in mAP10, where higher is better. LS-other represents test-other of LibriSpeech and AIS2 represents the AISHELL-2 test set.}
    \label{tab:audio_encoder}
\end{table}

\begin{table*}[t]
\centering
\begin{tabular}{r|lllll|lll|lll}
\toprule
& \multicolumn{5}{c|}{Sound} & \multicolumn{3}{c|}{Music} & \multicolumn{3}{c}{Speech}  \\
\multirow{1}{*}{Method} & ESC50 & FSD50K & US8K & CD & VS & GTZAN & NS & BO & SCV1 & SCV2 & FSC \\ \midrule
BLAT~\cite{xu2024blat} & 80.6 & 31.3 & 77.3 & 17.6$^\dagger$ & 53.9$^\dagger$ & 10.0$^\dagger$ & 9.03$^\dagger$ & 31.4$^\dagger$ & 3.9$^\dagger$ & 2.2$^\dagger$ & 0.4$^\dagger$ \\
MS-CLAP-2023~\cite{elizalde2024natural} & 88.2 & 40.3 & 75.0 & \textbf{29.7} & 69.2 & 58.4 & \underline{\textbf{47.9}} & 46.6 & 16.4$^\star$ & 2.5$^\dagger$ & 0.3$^\dagger$ \\
L-CLAP$_{\text{Speech-Music}}$~\cite{laionclap2023} & 89.3 & 20.2$^\dagger$ & 72.7$^\dagger$ & 20.7$^\dagger$ & 64.5$^\dagger$ & 52.3$^\dagger$ & 29.7$^\dagger$ & \textbf{57.2}$^\dagger$ & 3.8$^\dagger$ & 3.8$^\dagger$ & 0.4$^\dagger$ \\
L-CLAP~\cite{laionclap2023} & \textbf{91.0} & 21.5$^\dagger$ & 77.0 & 18.3$^\dagger$ & \textbf{79.3}$^\dagger$ & 47.4$^\dagger$ & 26.1$^\dagger$ & 40.2$^\dagger$ & 3.8$^\dagger$ & 4.1$^\dagger$ & 0.3$^\dagger$ \\
M2D-Clap~\cite{niizumi24_interspeech} & 75.5 & {40.8} & 72.4 & 17.7 & 42.3$^\dagger$ & \textbf{75.2} & 23.4 & 47.0$^\dagger$ & 3.0$^\dagger$ & 2.1$^\dagger$ & 0.4$^\dagger$ \\
\hline
GLAP & 88.8 & \underline{\textbf{40.9}} & \textbf{78.9} & 20.5 & 75.1 & 69.6 & {31.3} & 36.5 & \textbf{96.6} & \textbf{95.8} & \textbf{75.6} \\
\bottomrule
\end{tabular}
\caption{Zero-shot evaluation performance. Results marked with $^\dagger$ were obtained from a public checkpoint, while \underline{underlined} entries used the corresponding training dataset and are therefore not truly zero-shot. Entries with $^\star$ used the 10-class variant instead of the 30-class version employed in this work. Best results are bolded and higher is better.}
\label{tab:zeroshot_eval}
\end{table*}

\begin{table}[tb]
\resizebox{\columnwidth}{!}{
\begin{tabular}{r|llll|llll}
\toprule
\multirow{3}{*}{Method} & \multicolumn{4}{c|}{{AudioCaps}} & \multicolumn{4}{c}{{Clotho}} \\
& \multicolumn{2}{c}{Text-to-Audio} & \multicolumn{2}{c|}{Audio-to-Text} & \multicolumn{2}{c}{Text-to-Audio} & \multicolumn{2}{c}{Audio-to-Text}\\
& R@1 & R@10 & R@1 & R@10 & R@1 & R@10 & R@1 & R@10 \\
\midrule
BLAT~\cite{xu2024blat} & 33.3 & 82.4 & 40.4 & 85.7 &  12.3 & 46.1 & 13.9 & 48.2 \\
LClap-Large~\cite{laionclap2023} & 34.2 & 84.1 & 43.1 & 90.1  & 15.3 & 51.2 & 20.8 & 60.0 \\
% MSCLAP-2022~\cite{elizalde2022clap} & 33.5 & 80.2 & 47.8 & 90.7 &  16.2 & 51.4 & 23.6 & 60.3 \\
MSCLAP2023~\cite{elizalde2024natural} & 35.6 &  - & 42.5 & - & -  & 15.7 &  22.9 & - \\
Wavcaps-CNN14~\cite{mei2023wavcaps} & 34.7 & 82.5 & 44.7 & 86.2 &  21.2 & 59.4 & 25.9 & 65.8 \\
Wavcaps-HTSAT~\cite{mei2023wavcaps} & 39.7 & 86.1 & 51.7 & 90.6 &  20.2 & 58.8 & 26.5 & 67.3 \\
Auto-ACD~\cite{sun2024auto} & 39.5 & 85.4 & 53.7 & 91.7 &  15.3 & 52.1 & 17.7 & 52.6 \\
% T-CLAP~\cite{yuan2024tclap} & 39.7 & 86.9 & 49.8 & 91.9 &  17.3 & 53.6 & 21.8 & 57.4 \\
MLCLAP~\cite{yan24_interspeech} & 40.7 & {87.8} & 50.1 & 92.8 &  18.8 & 59.0 & 21.1 & 62.5 \\
% Cacophony~\cite{zhu2024cacophony} & 41.0 & 86.4 & {55.3} & 92.4 &  20.2 & 58.8 & 26.5 & 67.3 \\
SoundVECaps~\cite{yuan2024sound} & 41.2 & 85.3 & 53.3 & 93.0 & - &- &-& -\\
% CED-Pretrain & 40.4 & 87.1 & 55.7 & 90.8 & & 23.6 & 64.9 & 29.3 & 68 \\
\midrule
GLAP & {41.7} &86.1 &  54.4 & 91.1 &  19.4 & 58.3 & 21.8 & 61.5 \\
\bottomrule
\end{tabular}%
}
\caption{Sound event retrieval results compared to baselines.}
\label{tab:soundevent_results}
\end{table}

% \vspace{-3em}
\subsection{Sound retrieval results}
\label{ssec:retrival}

The performance of GLAP on the widely used AudioCaps and Clotho datasets for English sound-event retrieval is presented in \Cref{tab:soundevent_results}.
GLAP demonstrates strong results on both benchmarks, surpassing other methods in Text-to-Audio retrieval (R@1) on AudioCaps while maintaining competitive performance on Clotho.

\subsection{Music and speech retrieval results}
\label{ssec:music_speech_eval}

We assessed GLAP's retrieval capabilities using the Librispeech test-other dataset for English speech and the AISHELL-2 test set for unseen Chinese. 
For music, we used the MusicCaps dataset~\cite{agostinelli2023musiclm_musiccaps}. 
As shown in \Cref{tab:retreval_speech_music}, GLAP significantly outperforms previous methods in both music and speech retrieval. 
It achieves over 93\% on the English LibriSpeech test set and 98\% on the Chinese AISHELL-2 test set, demonstrating exceptional multilingual performance.

\subsection{Zero shot evaluation}
\label{ssec:zeroshot_eval}
Zero-shot results are shown in \Cref{tab:zeroshot_eval}. 
Our model performs similarly to other baselines for sound and music classification. 
However, it significantly outperforms them in keyword spotting tasks, achieving 95.8\%, 96.6\%, and 75.6\% accuracy on the SCV1, V2, and FSC datasets, respectively.
% The strong performance on the FSC dataset, which requires understanding entire sentences, demonstrates our model's superior ability to align with spoken content.

\vspace{-1.3em}
\subsection{Multilingual capabilities}
\label{ssec:multilingual}
GLAP's multilingual capabilities are evaluated using zero-shot tests on the Multilingual Spoken Words (MSW) Corpus~\cite{mazumder2021multilingual}.
\begin{figure}[!htb]
    \centering
    \includegraphics[width=\linewidth]{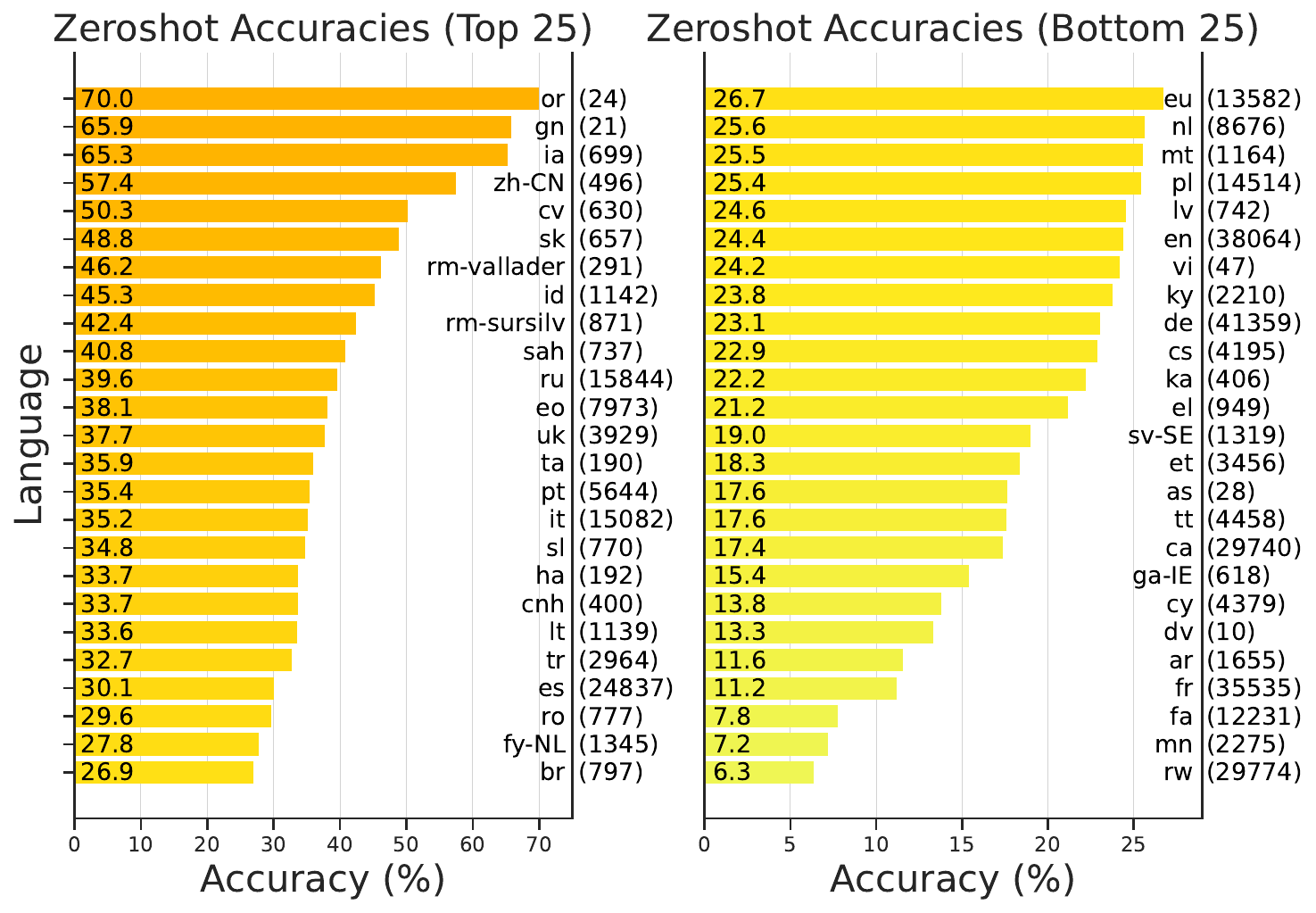}
    \caption{Multilingual zero-shot keyword spotting performance across 50 languages. The number of keywords (num) for each language is shown on the right.}
    \label{fig:msw_performance}
\end{figure}
Results in \Cref{fig:msw_performance} show strong performance across 50 languages. For example, Oriya and Guarani achieve the best accuracy at 70\% and 65.9\%, respectively. 
Notably, Chinese, with 496 keywords, reaches 57.4\% accuracy, while Russian, with 15,844 keywords, achieves 39.6\%.
We also assessed GLAP's multilingual sound and music understanding on the ESC-50, US8K, and GTZAN datasets.
As shown in \Cref{tab:ml_zeroshot_sound}, while performance decreases compared to the English baseline, GLAP remains effective for multilingual sound classification.

% The impressive performance in Russian is particularly noteworthy, as the model was trained only on Russian speech-text pairs from YODAS, not on music/sound pairs. This highlights GLAP's ability to effectively transfer multilingual text-based knowledge to the audio domain.

\begin{table}[tb]
\centering
\begin{tabular}{r|lllll}
\toprule
\multirow{2}{*}{Data} & \multicolumn{5}{c}{Language}  \\
  & {\color{xiaomi_gray}{En}} & De & zh-CN & Jp & Ru \\ 
\midrule
US8K     & {\color{xiaomi_gray} 78.9}  & 74.8 & 66.1 & 72.2   &  49.0  \\ 
ESC-50   & {\color{xiaomi_gray} 88.8} & 64.3 & 71.4 & 74.3 &  62.1   \\ 
GTZAN &  {\color{xiaomi_gray} 69.6} & 68.3 & 62.5 & 63.2 &  65.3 \\
\bottomrule
\end{tabular}
    \caption{GLAP's zero-shot evaluation for multilingual sound and music. Original labels ({\color{xiaomi_gray} {in gray}}) are translated into the target language using ChatGPT.}
    \label{tab:ml_zeroshot_sound}
\end{table}

\section{Conclusion}
\label{sec:conclusion}

We introduced GLAP, a versatile language-audio pretraining framework that enables multilingual and multi-domain modeling of both audio and text. 
To the best of our knowledge, it is the first \textit{single} system to integrate general audio and text embeddings into a unified contrastive framework.
GLAP demonstrates competitive performance on well-established benchmarks like AudioCaps and Clotho, while surpassing previous methods in music and speech retrieval tasks. 
Zero-shot evaluations show strong results for English sound and music tasks, extending effectively to other languages. 
Inference on the Multilingual Spoken Words dataset highlights robust multilingual capabilities beyond English.

\vfill\pagebreak

% \section{REFERENCES}
% \label{sec:refs}

\footnotesize

\enlargethispage{3\baselineskip}
\bibliographystyle{IEEEbib}
\bibliography{refs}

\end{document}